# Task allocation and site fidelity jointly influence foraging regulation in honey bee colonies

Thiago Mosqueiro[1], Chelsea Cook[2], Ramon Huerta[1], Jürgen Gadau[2], Brian Smith[2], Noa Pinter-Wollman[1,3]

[1] BioCircuits Institute, University of California San Diego, La Jolla, USA
[2] School of Life Sciences, Arizona State University, Tempe, USA
[3] Department of Ecology and Evolutionary Biology, University of California Los Angeles, Los Angeles, USA

## Abstract

Variation in behavior among group members often impacts collective outcomes. Individuals may vary both in the task that they perform and in the persistence with which they perform each task. Although both the distribution of individuals among tasks and differences among individuals in behavioral persistence can each impact collective behavior, we do not know if and how they jointly affect collective outcomes. Here we use a detailed computational model to examine the joint impact of colony-level distribution among tasks and behavioral persistence of individuals, specifically their fidelity to particular resource sites, on the collective tradeoff between exploring for new resources and exploiting familiar ones. We developed an agent-based model of foraging honey bees, parameterized by data from 5 colonies, in which we simulated scouts, who search the environment for new resources, and individuals who are recruited by the scouts to the newly found resources, i.e., recruits. We varied the persistence to return to a particular food source of both scouts and recruits and found that for each value of persistence there is a different optimal ratio of scouts to recruits that maximizes resource collection by the colony. Furthermore, changes to the persistence of scouts induced opposite effects from changes to the persistence of recruits on the collective foraging of the colony. The proportion of scouts that resulted in the most resources collected by the colony decreased as the persistence of recruits increased. However, this optimal proportion of scouts increased as the persistence of scouts increased. Thus, behavioral persistence and task participation can interact to impact a colony's collective behavior in orthogonal directions. Our work provides new insights and generates new hypotheses into how variation in behavior at both the individual and colony levels jointly impact the trade-off between exploring for new resources and exploiting familiar ones.

# Introduction

Group composition impacts the emergence of collective behaviors. Individuals that comprise a group vary both in which tasks they perform [1,2] and in how persistently they perform them, i.e. how many times they repeatedly perform a task [3,4]. The effect of allocation of workers to different tasks on the collective behavior of colonies has been studied extensively [5] with the underlying assumption that dividing the labor among group members will increase the overall efficiency of the group, as it does in human industrial societies [6]. However, variation among individuals in how persistently they perform a task is striking. This behavioral variation can undermine the efficiency that is often associated with task specialization [7,8] because individuals that are not persistent either do not perform a large proportion of the task or incur the costs of task switching [9,10]. Although recent work has begun to examine the impact of variation in individual persistence in performing a particular task on collective behaviors [11,12], we do not know how task allocation and variation in persistence interact to impact collective outcomes.

Behavioral persistence has now been documented extensively throughout the animal kingdom [3] including in social insects [4]. Some ant workers are more persistent in performing a certain task than others [13], and honey bee workers vary in how persistently active they are [14,15]. Behavioral persistence can impact how individuals in a group interact with one another and therefore affect the collective behaviors that emerge from these interactions [12,16]. A growing understanding of the mechanisms that underlie behavioral persistence is paving a path for understanding how variation in behavioral persistence affects collective outcomes. For example, the decision of a honey bee to leave the hive and start foraging is influenced by the bee's genotype [17–22]. Furthermore, genetic variation underlies individual differences in learning abilities, which might influence the likelihood of a bee to make certain types of foraging decisions, such as staying in or leaving a resource patch [23–27].

Honey bees exhibit variation in foraging behaviors at both the worker and colony levels [28,29]. Understanding the mechanisms that underlie honey bee foraging decisions is especially important because of their economic importance for honey production and crop pollination [30,31]. Consistent behavioral variation across workers within honey bee colonies has potential fitness consequences [28]. Although the regulation of foraging behavior in honey bees has been studied for a long time [32], and much is known, for example, about how foragers respond to resource availability [21,33], we still do not know what mechanisms may underlie variation among colonies in collective foraging.

Many tasks in honey bee colonies are related to foraging. For example, some foragers collect pollen while others specialize on collecting nectar [32,34], and an animal's genotype influences a bias for one or the other [35–37]. Nectar foragers further vary in their propensity to leave the nest to find new food. Experienced foragers that spontaneously leave the hive to explore the environment are referred to as 'scouts' or 'primary searchers' [38–41]. When these scouts return to the hive, they recruit other foragers to the food patches they found and these bees are referred to as 'recruits'. Scouts communicate to recruits the direction, distance, and quality of newly found resources using the waggle dance [32], thus reducing waste of energy spent when searching for food over both long and short time scales [42,43], and dangers, such as predation [33,44,45]. Although exploration of the environment for new food sources is a task exclusive to scouts, they can contribute to the exploitation of resources, alongside the recruits, through repeated visits to the same source [21,46]. We define persistence of a forager as the average number of repeated visits it performs to each particular resource, regardless of whether it is a scout or a recruit. Thus, both scouts and recruits with lower persistence can contribute to a colony's exploration of the environment because low-persistence scouts will travel to different resources and low-persistence recruits will stop foraging quickly and become available to be recruited to new locations. High persistence of both scouts and recruits can contribute to the colony's exploitation of resources through repeated visits to a profitable source but can also hinder the efficiency of collective foraging if other, more profitable resources are available. Honey bees choose between exploring for new resources or exploiting familiar ones based on colony [40] and individual information [47,48]. Thus, the tradeoff between exploration and exploitation can be adjusted both at the colony level, through allocation of foragers to either scouts or recruits, and at the individual level, through variation in the persistence of visits to a known food source.

Although the tradeoff between exploration and exploitation has been previously examined in honey bees by addressing the differences between scouts and recruits [21,49], the role of behavioral persistence in visiting the same resource, i.e., site fidelity, has thus far been overlooked. Because foraging is energetically costly [47,50], greater persistence does not always translate into greater efficiency. To examine the joint role of task allocation and behavioral persistence in the regulation of foraging by honey bees we considered how the ratio between scouts and recruits along with the persistence of returning to a particular resource jointly affect the collective resource acquisition by a colony. Specifically, we examine how behavioral persistence of (i) the entire colony, (ii) scouts, or (iii) recruits affects collective foraging when different proportions of foragers are allocated to either scouting or being recruited. Our findings provide new and realistic insights on how behavioral variation at more than one level of organization impacts collective outcomes.

# Materials and methods

## Agent-Based Model

To examine the joint impact of task allocation and behavioral persistence on collective behavior we developed a spatially-explicit Agent-Based Model. Simulated honey bees foraged in an open, continuous, 2D space. The hive was set at the origin of the space and three unlimited resource patches were uniformly distributed around it at a fixed distance of 15m from the hive and equal distances between neighboring sites. We simulated two types of foragers: scouts and recruits, which varied in their flight patterns as detailed below. To determine the effects of behavioral persistence on colony outcomes we examined the proportion of scouts that lead to the maximum amount of resources collected by the colony under different regimes of behavioral persistence. A description using Overview, Design concepts, and Details (ODD) protocol [51] and the source code of our model can be found on Github [52].

## Flight dynamics

Flight dynamics of all foragers were modeled as a random walk with drift [53,54]. At the beginning of each simulation ($t = 0$), the position of each bee $i$ was $\boldsymbol{x}_i(0) = (0,0)$, i.e. all bees were at the hive. Each bee was assigned a different drifting vector $\boldsymbol{v}_i$, which determined its flight direction when leaving the hive and its flight pattern is described as

$$d\boldsymbol{x}_i(t) = \boldsymbol{v}_i \, dt + \sigma_i \, d\boldsymbol{W}_t, \tag{1}$$

where $\sigma_i \, d\boldsymbol{W}_t$ is the random contribution to the distance and angle a bee moved. This term has a normal distribution with a mean of zero and variance of $\sigma_i$, thus closely resembling a diffusion process [55]. Specifically, $1/\sigma_i$ measures the precision of the flight. Because $E[d\boldsymbol{W}_t] = \boldsymbol{0}$, the average velocity of the $i$-th bee was $\boldsymbol{v}_i$, and its magnitude $v_i = |\boldsymbol{v}_i|$ defined the average flight velocity. The stochastic dynamics in Equation (1) produces slight variation among bees in their flight patterns to avoid an unrealistic scenario in which bees take a straight line between two points. Using the Euler–Maruyama method [56], equation (1) can be solved numerically using

$$x_i(t + \Delta t) = v_i \Delta t + \sqrt{\Delta t}\, \sigma_i W_t + x_i(t) = v_i \Delta t + \tilde{\sigma}_i W_t + x_i(t), \tag{2}$$

with $\Delta t$ being a fixed time step, and $\tilde{\sigma}_i = \sqrt{\Delta t}\, \sigma_i$. At the beginning of each simulation (t=0) scouts left the hive, with drifting vectors $v_i$ assigned from a uniform distribution, and continued flying until they found a resource. Once a scout detected a resource, it returned to the hive to recruit other foragers, referred to as "recruits". Scouts and recruits differed in the precision of their flight: $\tilde{\sigma}_i$ of scouts was larger ($\tilde{\sigma}_i = 5$) than that of recruits resulting in flight paths that covered a larger area than recruits (Figure 1). The dispersion of recruited bees ($\tilde{\sigma}_i = 2$) was fitted using data from experiments with feeders positioned at distances varying from meters to kilometers [32]. To differentiate between the flight patterns of bees that are exploring the environment and those that are exploiting a resource patch, are familiar with its location, and are therefore faster and more precise, we assigned $v_i = 1$ to scouts, and $v_i = 1.5$ to recruits, following [57]. Foragers that reached the limit of the simulated area were set back to the hive instantly to start foraging again.

During recruitment, scouts communicated the location and distance of the newly found resource. The recruiting scout remained at the hive for 1min (approximately 50 time steps in the numeric simulations) to simulate the time it would take to recruit foragers using the waggle dance [32]. During this period, an average of five randomly selected recruits left the hive in the direction of the resource. Recruiting on average 1, 5, or 10 foragers by each scout did not qualitatively change the results of our simulations. For simplicity, only the recruitment by scouts is considered here, and we examine the effect of adding recruitment by recruits in the Supplemental Material (Figures S4 and S5).

Each of the newly recruited bees left the hive with their drifting vectors pointing exactly towards the location reported by the recruiting scout. The direction of this drifting vector is the deterministic portion of the flight dynamics (see $v_i\, dt$ in equation 1), which is accompanied by a stochastic contribution from $\sigma_i\, dW_t$. Recruited bees exploited the first resource they found during their trips. The dispersion of recruited bees (σ=2) was fitted using data from experiments with feeders [32]. Because the stochastic element of the flight of a recruited bee is very small compared to the size of the resource patches in our simulations, bees always exploited the same resource patch that was reported to them. The effect of communicating the distance to the source was modeled by slightly changing the dynamics in equation (1) to

$$dx_i(t) = v_i\, \alpha(|x_r - x_i(t)|)\, dt + \sigma_i\, dW_t, \tag{3}$$

where $\alpha(x)$ is any function that goes to zero when $x \to 0$ and $\boldsymbol{x}_r$ is the location of the resource reported. This turned the flight dynamics into a purely random walk (i.e., without bias) near the location of the reported resource. For simplicity, we used a Heaviside function that removed all bias in the flight dynamics when the forager was less than 2m from the resource:

$$\alpha(\boldsymbol{x}_r - \boldsymbol{x}_i(t)) = \begin{cases} 1, & \text{if } |\boldsymbol{x}_r - \boldsymbol{x}_i(t)| - 2 \text{ ;} \\ 0, & \text{otherwise.} \end{cases} \qquad (4)$$

During our simulations, scouts and recruits obtained resources for the colony. Upon arriving at a resource, foragers (both scouts and recruits) returned to the hive in a straight line, with constant velocity $v_i$, carrying one resource unit, equivalent to $1.0 \pm 0.3\ \mu\ell$ [32]. If a forager reached the boundaries of the area considered in the simulation, it was re-assigned to the hive, without bringing food, to begin foraging again. Each forager, scout or recruit, was assigned a persistence value $\pi_i$, defined as the number of consecutive trips it performed to each resource location. If the persistence of a scout was greater than 1, its $v_i$ and $\sigma_i$ after the first trip were set to those of recruits and its flight dynamics was adjusted to follow equation (2). Scouts that completed $\pi_i$ trips to the same location randomly changed their drifting vector and began scouting again. Recruits that completed $\pi_i$ trips, remained at the hive until they were recruited again.

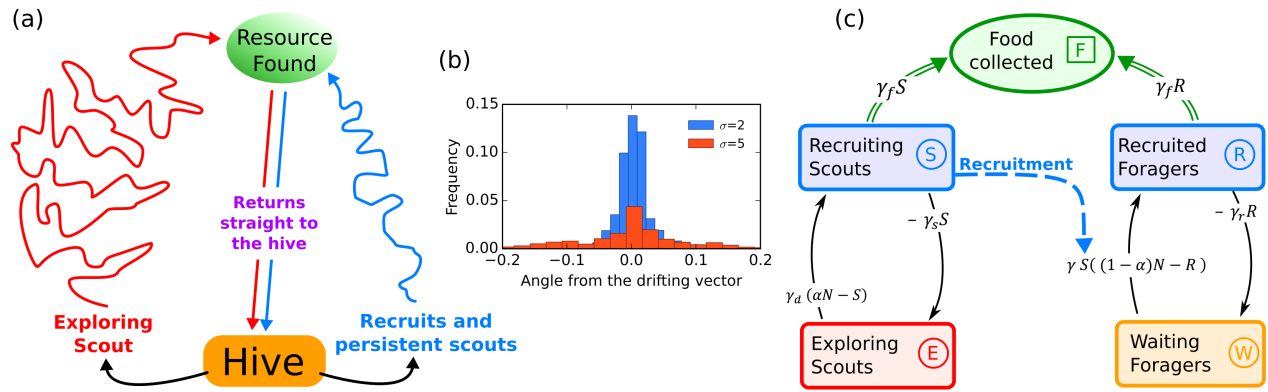

**Figure 1.** Flight dynamics of scouts and recruits. **(a)** Scouts left the hive at the beginning of the simulation and once they found a resource, they recruited other foragers, referred to as 'recruits'. **(b)** Variance of the scouts' deviations from a straight path on outgoing trips ($\tilde{\sigma} = 5$, red) was larger than that of the recruits and persistent scouts ($\tilde{\sigma} = 2$, blue) resulting in greater spatial dispersion. **(c)** System Dynamics approach based on a compartmental model, with square boxes representing the states of foragers and the green circle representing the amount of food retrieved by all foragers. Black arrows are state-transition rates (see equations 6 and 7); the blue dashed arrow represents the recruitment of foragers by scouts; the green double arrows represents foragers delivering food to the hive.

## Collective outcomes

To examine the impact of colony composition on collective foraging success, we simulated colonies with different ratios between scouts and recruits. Simulated colonies consisted of 300 foragers that were allowed to forage for 7h in an area equivalent to 1.3km². These values were selected based on empirical data on honey bee foraging [32]. Because each simulation reflected just one day of foraging, we assumed that resources were never depleted during a simulation and that the ratio between scouts and recruits was fixed.

The colony-level outcome was measured as the total amount of resources retrieved by all the bees in the colony. For each simulation j that we ran, we recorded the resources $f_j(t)$ collected over time. Because of the stochastic nature of our simulations, the amount of resources collected at each time point over all our $n$ simulations followed a bell-shaped distribution with a variance $V$. To ensure that all conditions tested (i.e., proportion of scouts and various persistence values) produced the same 90% confidence interval $w$ for the estimation of the average amount of resource collected (see shaded area in Figures 3a,b) we used the central limit theorem to set the number of simulation runs to $n = 4V^2/w$. Because the mean of the total amount of resources collected was on the order of thousands of microliters, we set $w = 50\mu\ell$, resulting in $n$ of approximately 120. We estimated the average amount of resources collected at every time point in all $n$ simulation runs as $f(t) = E[f_j(t)]$ (see lines in Figures 3a,b).

## System Dynamics Model

To complement our understanding of how behavioral persistence and recruitment by scouts in the Agent-Based Model combine to result in complex outcomes, we used a coarse-grained formalism based on ordinary differential equations that describe the system's dynamics (Figure 1c), similar to [58]. We consider the following dynamical variables: E(t) - the number of scouts exploring the environment; S(t) - the number of scouts that are bringing food back to the hive; R(t) - activated recruits; and W(t) - potential recruits waiting inside the hive. Let N be the number of foragers in the colony, then $\alpha N$ is the total number of scouts and $(1 - \alpha)N$ is the total number of recruits. Thus, $E(t) = \alpha N - S(t)$ and $W(t) = (1 - \alpha)N - R(t)$. Because S(t) and R(t) represent the total number of foragers collecting food at any given time, we refer to them as *active foragers*. If we define $\gamma$ as the rate at which active scouts S(t) recruit inactive recruits W(t), then the increase in the number of active recruits is described by

$\gamma SW = \gamma S ((1 - \alpha)N - R)$. A simple model describing the rate of change in number of scouts and number of recruits can be defined by two differential equations:

$$\frac{dS}{dt} = \gamma_d (\alpha N - S) - \gamma_s S, \tag{5a}$$

$$\frac{dR}{dt} = \gamma S((1 - \alpha)N - R) - \gamma_r R, \tag{5b}$$

where $\gamma_d$ is the rate at which scouts find a new resource and start exploiting it; $\gamma_s$ is the rate at which these scouts stop collecting food and resume exploring for new resources; and $\gamma_r$ is the rate at which the recruited foragers stop collecting food and begin waiting to be recruited again. Finally, the cumulative amount of food collected by active foragers F(t) can be formulated as

$$\frac{dF}{dt} = \gamma_f (S + R), \tag{6}$$

with $\gamma_f$ being the rate at which bees collect food while exploiting a particular resource.

In this compartmental model, behavioral persistence, in the form of repeated visits to a particular resource site, is defined according to the rates at which foragers stop exploiting particular resources. Both $1/\gamma_s$ and $1/\gamma_r$ represent the characteristic durations of exploiting a particular resource by scouts or recruits. Dividing these characteristic durations by the average time interval $<\Delta>$ between each visit to the feeder (which was experimentally evaluated as described below in the section 'Behavioral Experiments and Parameter Estimation') gives the average number of visits to one resource. Thus, to link the rates at which foragers stop exploiting a particular resource with the persistence parameter in the Agent-Based Model, we define

$$\gamma_s = \frac{\overline{\gamma_s}}{\pi^s <\Delta>}, \tag{7a}$$

$$\gamma_r = \frac{\overline{\gamma_r}}{\pi^r <\Delta>}. \tag{7b}$$

Defining the relationship between $\gamma_s$ and $\gamma_r$ and persistence, as simulated in the Agent-Based Model, allows us to analyze the compartmental model without having to fit a different value of $\gamma_{s,r}$ for each $\pi^{s,r}$, reducing the complexity of our compartmental model. The parameters $\overline{\gamma_s}$, $\overline{\gamma_r}$, $\gamma_f$ and $\gamma_d$ were fitted using simulation data from the Agent-Based Model.

# Behavioral Experiments and Parameter Estimation

To assess persistence empirically we observed the visitation of 323 honey bee (*Apis mellifera* L.) foragers from five different colonies during the winter (between February 3[rd] and 26[th], 2016). Each colony was tested on a different day and was presented with two feeders, each containing 1M sucrose solution on which the foragers fed *ad lib*. We trained bees to find feeders, positioned at 3, 5 or 10m from the hive, one day before the experiments began, following [18] and comparable to other studies that examine 20m [59]. During the time of our experiments there were few naturally blooming plants and our feeders were very attractive to the bees. We marked workers for individual identification using water-based acrylic paint markers (Montana) and recorded the time at which each bee visited a feeder using the software EventLog [60]. We recorded 1307 trips. Work with invertebrates does not require the ethics committee approval and all field work was conducted on university property. All collected data is publicly available [61].

We estimated the values for the parameters in our model based on the empirical observations. Interestingly, all bees exhibited the same rate of visits to the feeders (Figure 2a), which was $0.4 \pm 0.2$ visits per minute (Figure 2b). This visitation rate allowed us to set the model parameter $v_i$ for flight velocity to a constant value for all foragers after their first visit at a resource. The empirical distribution of intervals between consecutive visits to the feeder (Figure 2b) informed the visitation interval of our model. The observed average visitation interval $<t>$ was linearly related to the distance $d$ between hive and feeder: $<\Delta> = \alpha d + \beta$, with α = 2.3 $\pm$ 0.3 and β = 0.28 $\pm$ 0.05. Finally, the observed distribution of persistence was geometric or negative binomial (Figure 2c), with an average of $\pm$ 90% CI = 6.1 $\pm$ 0.3. This means that making the decision to stop exploiting a particular patch had a probability of 16% based on the value of the lambda parameter of a geometric distribution that was fit to the data. Because the largest number of observed return visits by a single bee was 22, we restricted our persistence parameter $\pi_i$ to range between 1 and 30.

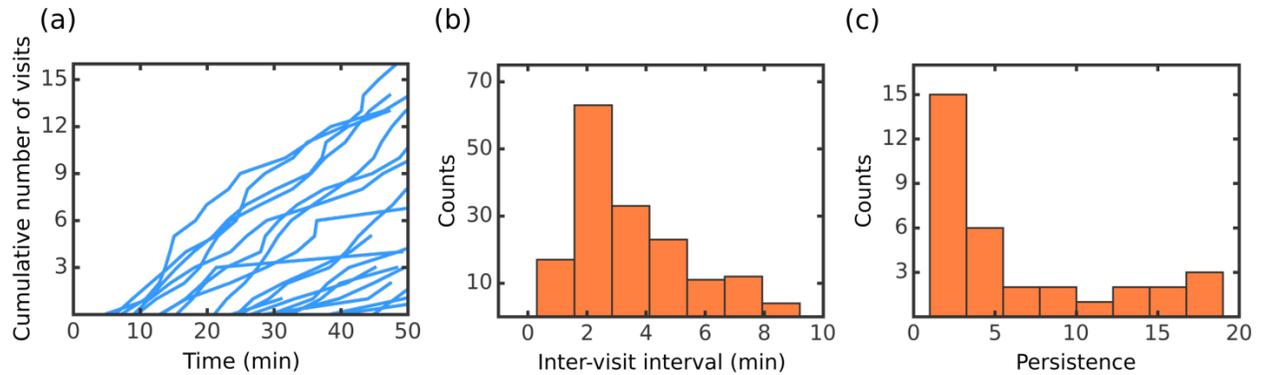

**Figure 2.** Empirical results of 206 foraging trips performed on one day by 33 different honey bee foragers from one representative colony of the five colonies we tested. The feeder was positioned 10m from the hive. (a) Number of visits over time. Each line represents one bee and t=0 reflects the first bee's first visit to the feeder. (b) Distribution of intervals between consecutive visits to a single feeder. (c) Distribution of persistence, i.e., the number of return visits by each bee to one of two feeders.

# Results

The proportion of scouts that maximized the amount of resources a colony collected by the end of the simulation, referred to as the 'optimal proportion of scouts', changed with the persistence of visiting a resource. The amount of resources collected in all simulations increased over time (Figures 3a,b). The total amount of resources collected at the end of the simulation was different among the various proportions of scouts. When both scouts and recruits lacked persistence, i.e., each bee made only a single trip to the feeder ($\pi = 1$), more resources were collected as the proportion of scouts increased (Figure 3a). However, as the persistence of all foragers increased from $\pi = 1$ to $\pi = 20$, a greater proportion of scouts in a colony did not necessarily result in more resources collected. For example, when persistence was set at $\pi = 20$, colonies with 50% scouts outperformed colonies with 90% scouts (Figure 3b,c). For each persistence value $\pi$ we found the optimal proportion of scouts, i.e., the proportion of scouts that resulted in the most resources collected by the end of the simulation (after 7h of foraging) (Figure 3c). This optimal proportion of scouts decreased with persistence and saturated after $\pi > 20$ (Figure 3d). However, the absolute amount of resources collected per colony continued to grow when persistence increased beyond 20 visits per individual ($\pi > 20$) (Figure 3e). Changing the number of resource patches impacted the total amount of resources collected by the colony, but the optimal proportion of scouts still decreased with the persistence of the colony (Figure S6).

The System Dynamics model allows us to further evaluate the processes that determine the optimal proportion of scouts using the stable solutions for scouts (S) and recruits (R),

$$S^* = S(\infty) = \alpha N \left( \frac{\gamma_d}{\gamma_d + \gamma_s} \right), \tag{8a}$$

$$R^* = R(\infty) = (1 - \alpha)N \left( \frac{1}{1 + \frac{\gamma_r (\gamma_d + \gamma_s)}{\gamma \gamma_d \alpha N}} \right). \tag{8b}$$

The expressions inside the parenthesis in Equations 8a and 8b represent respectively the proportions of scouts and recruits that become active after a long time, *i.e.*, asymptotically. These solutions reveal that the optimal proportion of active scouts is determined solely by the ratio between the rate at which scouts discovering new resource sites, $\gamma_d$, and the rate at which they abandon them, $\gamma_s$. For a fixed rate of discovery, $\gamma_d$, the number of active scouts increases almost linearly with the persistence of scouts, saturating for large values of persistence, i.e. when $\gamma_s \to 0$. However, the number of active recruits, however, does not depend directly on the persistence of scouts, but on the number of scouts, $\alpha N$, and the rate of recruitment, $\gamma$. From Equation 7, the amount of food collected, F(t), grows asymptotically at a fixed rate,

$$\frac{dF}{dt} = \gamma_f (S^* + R^*) = \gamma_f N \chi_2 \alpha \left[ 1 + \chi_1 \frac{1 - \alpha}{1 + \alpha \chi_1 \chi_2} \right], \tag{9}$$

where $\chi_1 = \gamma N / \gamma_r$ measures the trade-off between recruitment and the persistence of recruits; and $\chi_2 = \gamma_d / (\gamma_d + \gamma_s)$ is the ratio between the rate of discovering new resource sites $\gamma_d$ and the rate of abandoning a site $\gamma_s$ (same expression as in 8a). If there are no scouts, $\alpha=0$, then no food is collected, which agrees with the Agent-Based Model (Figure 3c). Because the rate $\gamma_d$ at which new resources are discovered is constant in our model, the amount of food collected, F(t), always grows and does not present a stable solution. However, the asymptotic speed at which F(t) grows, shown in Equation 9, changes with the proportion of scouts in the colony, $\alpha$. Thus, for long times, the amount of food collected, F(t), grows linearly, and comparing the rate of increase among different persistence values is equivalent to comparing the relative values of F(t) at a fixed time point t, as in Figures 3c, d. To simplify the dependence of the rate of increase of F(t) on the proportion of scouts, $\alpha$, in Equation 9, we use the Taylor expansion up to second order in $\alpha$:

$$\gamma_f (S^* + R^*) = \gamma_f N \chi_2 [\alpha(1 + \chi_1) - \alpha^2 \chi_1 (\chi_1 \chi_2 + 1)] + \mathcal{O}(\alpha^3), \tag{10}$$

with $\mathcal{O}$ being the "big O" notation, i.e. refers to the remaining terms that are polynomials in $\alpha$ of order 3 or higher, and have a small contribution to Equation 10 because $0 < \alpha \leq 1$. Thus, the asymptotic rate of resource collection is a concave function whose maximum depends on α, the proportion of scouts in the colony, in accordance with the results from our spatially-explicit Agent-Based Model (Figure 3c). The optimal proportions of scouts predicted by the System Dynamics agree perfectly with the results of the Agent-Based Model (lines in Figures 3c, d). However, the curvature of the amount of resources collected in relation to the proportion of scouts slightly differs between the System Dynamics and the Agent-Based models (Figure S7).

Changes in the persistence of scouts had the opposite effect from changes in the persistence of recruits on the proportion of scouts that maximized collective resource collection. In the Agent-Based Model, while the optimal proportion of scouts decreased with the persistence of recruits $\pi^r$ (Figure 4a), this proportion increased with persistence of scouts $\pi^s$ (Figure 4b). This opposite dependence of the optimal proportion of scouts on $\pi^s$ and $\pi^r$ was observed for a wide range of both scout and recruit persistence values (Figures 4c, d). Our System Dynamics model also reproduces this dependence (see lines in Figures 5). The combined scout-recruit persistence with the best collective outcome, i.e., greatest amount of resources collected, resulted from the largest persistence values of both scouts and recruits (Figure 4e) when approximately 60% of the foragers were scouts (Figure 4f). The opposing dependence of the optimal proportion of scouts on scout and recruit persistence is captured by our System Dynamics (Figure 5), through the relationship between $\gamma_s$ and $\gamma_r$ in Equation 10. Interestingly, changes in the persistence of recruits resulted in a 50% change to the optimal proportion of scouts, whereas changes in the persistence of scouts resulted in only a 25% change in this proportion (Figure 6).

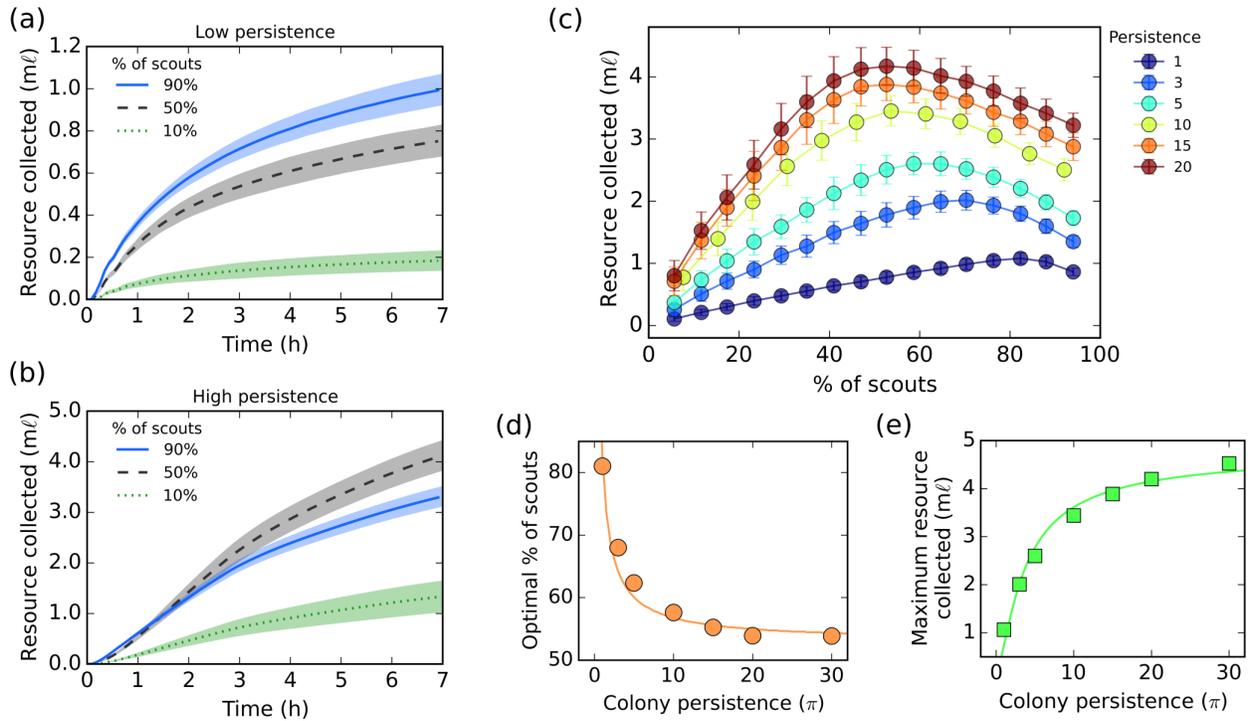

**Figure 3.** The relationship between colony persistence $\pi$ and the proportion of scouts affects the amount of resources collected by a colony. The amount of resources collected over time by a simulated colony in which all foragers have either (a) low persistence ($\pi = 1$) or (b) high persistence ($\pi = 20$) for three different proportions of scouts. Shaded areas represent 1.5 standard deviation. (c) Total amount of resource collected throughout the entire simulation as a function of the proportion of scouts in the colony for different values of persistence of all foragers ($\pi$). Bars are the standard deviation across all simulation runs. (d) Optimal proportion of scouts plateaus near 50% as $\pi$ increases. Points are the results from our Agent-Based Model and the line is the result from the Systems Dynamics approach (Equation 10). (e) Maximum amount of resources collected scales sublinearly with $\pi$. Points are the results from our Agent-Based Model and the line is the result from the Systems Dynamics approach (Equation 10).

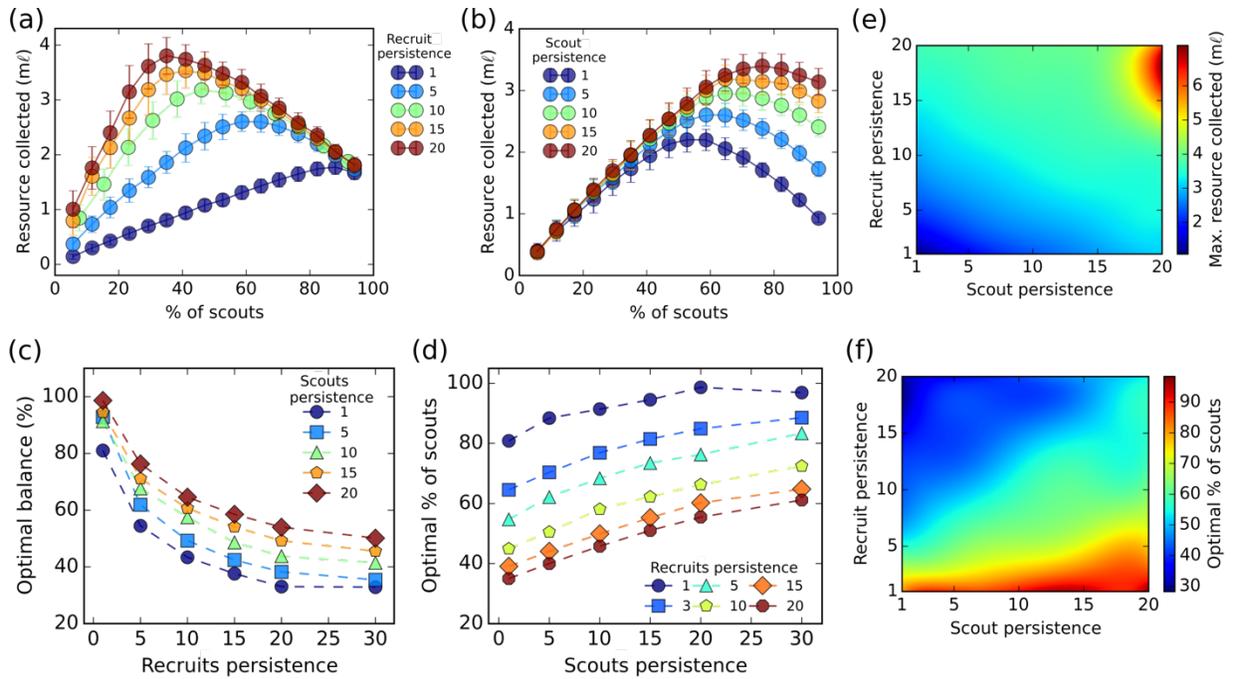

**Figure 4.** Differences in collective foraging due to the persistence of either scouts ($\pi^s$) or recruits ($\pi^r$) in the Agent-Based Model. Total amount of resources collected by a colony as a function of the proportion of scouts when (a) the persistence of scouts is set to $\pi^s = 5$ for the following values of persistence for recruits: $\pi^r$ =1,5,10,15,20; and (b) the persistence of recruits is set to $\pi^r = 5$ for the following values of persistence of scouts: $\pi^s$ =1,5,10,15,20. Bars are the standard deviation across all simulation runs. Proportion of scouts that resulted in maximal amount of resource collected as a function of (c) recruit persistence for different values of fixed scout persistence $\pi^s$; and (d) scout persistence for different values of fixed recruit persistence $\pi^r$. (e) Heat map of the maximum amount of resources collected for different values of scout $\pi^s$ and recruit $\pi^r$ persistence jointly. (f) Heat map of the proportion of scouts that led to the maximum amount of resources collected for different values of scout $\pi^s$ and recruit $\pi^r$ persistence jointly.

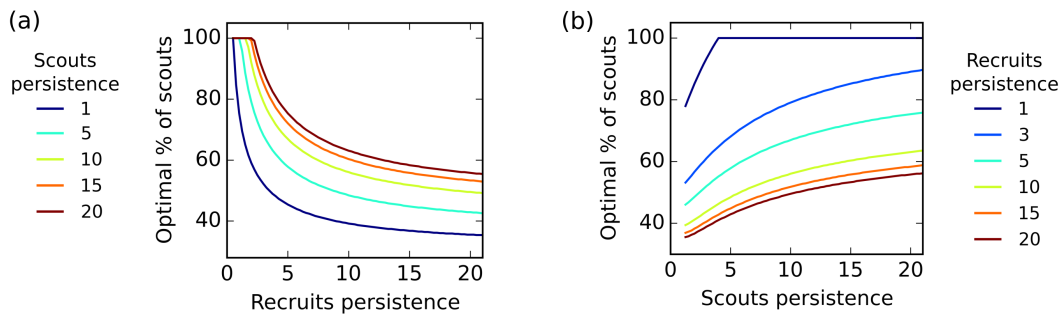

**Figure 5.** The Systems Dynamics approach captures the opposing effects of scout and recruit persistence on the optimal proportion of scouts. (a) Change in optimal percent of scouts due to changing the persistence of scouts. (b) Change in optimal percent of scouts due to changing the persistence of recruits.

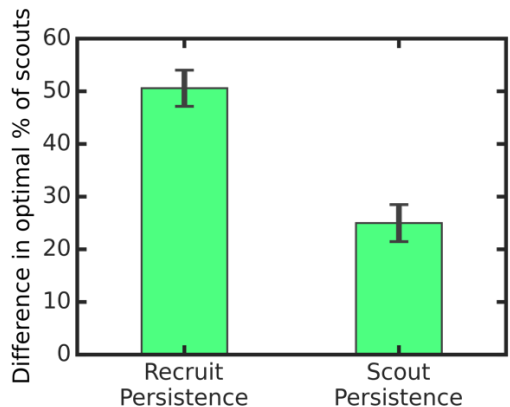

**Figure 6.** The effect of increasing recruit persistence on the proportion of scouts that resulted in an optimal amount of resource collected was double that of the effect of increasing scout persistence. Bars are the standard deviation across all persistence values.

# Discussion

Social groups constantly adjust their collective behavior to changes in their surroundings. However, an understanding of how these adjustments emerge is still scant. Our models show that both colony-level composition, i.e., the ratio between scouts and recruits, and individual-level traits, such as the persistence of foragers, interact to impact collective foraging. We found that the balance between the proportion of bees scouting and behavioral persistence allows a colony to acquire more resources and allocate fewer individuals to the potentially costly activity of scouting. Scouts may expend considerable energy flying around in search for new resources, and they can be preyed upon or potentially lose their way home [33]. In our simulations, colonies with high persistence, $\pi = 20$, collected almost five times more resources than those with low persistence, $\pi = 1$ (Figure 3c). The tradeoff between exploring for new resources and exploiting known ones resulted in a different optimal proportion of scouts for each value of persistence (Figure 3). As persistence increased, the proportion of scouts required for collecting the maximal amount of resources decreased to a minimum near 50% (Figure 3b-c), because exploiting known resources required fewer scouts to find new resources. Previous studies estimated that the percentage of scouts in honeybee colonies is between 5% and 35% [38]. These numbers are slightly lower than the optimal proportions we found here and this difference can be explained by the parameter choice for the number of bees that respond to a waggle dance (Figure S3). Although colonies with higher recruitment rates required fewer scouts to achieve their optimal foraging performance, the relationship between the optimal proportion of scouts and persistence did not change (compare Figures 3 and 4 with Figure S3). Changing the number of foragers (from 100 to 1500) did not qualitatively change how persistence and colony composition interacted to achieve optimal resource collection (Figure S2), although, in agreement with previous modeling efforts [57], larger colonies did induce faster collection of resources. Lastly, the effect of including recruitment by recruits on the optimal proportion of scouts was the same as that of increasing the number of recruited foragers by scouts per waggle dance (Figures S4 and S5).

Changing the persistence of scouts had a different impact on collective foraging than changing the persistence of recruits. We found that increasing the persistence of recruits resulted in a decrease in the proportion of scouts required for collecting the maximal amount of resources. In contrast, increasing the persistence of scouts resulted in an increase in the proportion of scouts required for collecting the maximal amount of resources (Figures 4a-d). This result suggests that the persistence of recruits was the predominant factor impacting the optimal proportion of scouts when varying the persistence of all foragers (Figure 3). Indeed, the effect of the persistence of recruits on the proportion of scouts that resulted in an optimal collective outcome was double that of the impact of persistence of scouts (Figure 6). Because recruits spend much time inside the hive, their persistence may change in response to information about the amount of resource stocks in the hive [41,62]. Furthermore, recruits may acquire information from several scouts that are returning from different locations and decide which one to follow and how many trips to make to each location, depending on their relative quality [32,63,64]. If the persistence of recruits is flexible and is determined by integrating information about resources inside and outside the hive, the substantial impact of their persistence on collective foraging that we found suggests that recruits may be the ones driving the adjustment of the colony's exploration-exploitation strategy in response to both external and internal conditions. However, if behavioral persistence is not a flexible trait, perhaps because it is regulated by genetic or epigenetic/developmental factors [17–20,65], our simulations show that a colony can compensate for having highly persistent scouts by allocating more foragers to the scouting task. Interestingly, colonies with comparable persistence for scouts or recruits collected almost the same amount of resources (compare curves with same color in Figures 4a-b) but the optimal proportion of scouts required to achieve the maximal amount of resource collection differed between the two cases.

Learning the location of a resource did not affect the relationship between persistence and the proportion of scouts. In our simulations, bees communicated the location of newly discovered resources, which is known to increase resource collection in patchy environments [57,66]. Our incorporation of behavioral persistence further enhanced this positive effect of communication by effectively simulating 'learning' of the target location. Return flights of scouts to a particular resource became more precise than their initial flight during which they located the resource (Figure 1b). Interestingly, when this learning was removed, i.e., flights did not become more precise, the relationship between the optimal number of scouts and persistence was unchanged but the rate of resource collection substantially decreased (Figure S1). Thus, when repeatedly returning to the same location does not increase collection efficiency, the total benefits are reduced, but the collective dynamics which dictate the optimal proportion of scouts are unchanged. It would be interesting to further investigate the effect of increasing collection efficiency on collective dynamics in primitively social bees that exhibit division of labor but spatial information is not shared during recruitment e. g., bumble bees [43], or halictine bees in which there are no known mechanisms of recruitment [67].

The spatial and temporal abundance of resources can substantially impact foraging behavior [26,57,66,68]. Indeed, during the development of our model we found that increasing the number of resource patches caused the total amount of resources collected by the colony to increase for all proportions of scouts, and the optimal proportion of scouts to decrease with the number of patches (Figure S6). This finding is consistent with a model of collective foraging in ants [69] which also found that the optimal proportion of scouts is inversely related to the amount of resources in the environment. To examine the relationship between the proportion of scouts in a colony and behavioral persistence, without the confounding effects of resource distribution, we focused only on one spatial setting in our final model. The simulations we present have biological significance for foraging in patchy resources that cannot be depleted in a single day, such as blooming trees or meadows.

In conclusion, we showed that both colony-level composition and individual-level traits interact to impact collective outcomes. Therefore, to fully understand the tradeoff between exploring for new resources and exploiting familiar ones, we need to both (i) uncover the mechanisms that underlie behavioral persistence and task allocation, and (ii) determine the time scales on which these processes act in different species that live in different environments. Our model can thus be used to generate hypotheses for further empirical work on the regulation of collective behavior and its response to various environmental conditions.

# Data, code and materials

- All data collected on behavioral persistence is publicly available on FigShare [61]: https://figshare.com/s/71248965ef012e412c69
- Details and source code of our simulations are publicly available on Github [52]: http://github.com/VandroiyLabs/ABBAS

# Acknowledgements

We would like to thank the social insect research group at ASU for helpful comments and Dr. Byron Van Nest for comments on a previous version of this manuscript.

# Funding

Funding for this work was generously provided by the NIH grant R01GM113967 to BHS, NPW, JG and RH. TM acknowledges support from CNPq grant 234817/2014-3.

# Competing Interests

The authors declare no competing interests.

# Authors' Contributions

TM and NPW designed the simulations; TM and RH designed the ODEs; and TM performed all the computations. CK collected the data. TM, CK, NPW and RH analyzed the data. All authors participated in writing the paper and gave final approval for publication.

# ELECTRONIC SUPPLEMENTARY MATERIAL

## Increasing flight precision in response to finding a resource

Removing the increase in flight precision after detecting a new resource did not change our main findings. The increase in flight precision that we included in our model reflects the communication of information about distance and direction between scouts and recruits [1]. The proportion of scouts that led to an optimal amount of resources collected decreased with persistence in the presence and absence of increasing flight precision (Figure S1a). However, the total amount of resources that a colony collected when there was an increase in flight precision after finding a resource was substantially larger than without this increase (Figure S1b). In both cases, the maximum amount of resources collected can be approximated by a saturating exponential model, $A_0(1 - e^{\alpha \pi})$, with $\pi$ being persistence, and $A_0$ and $\alpha$ fitted using the simulated results (see lines in Figure S1b).

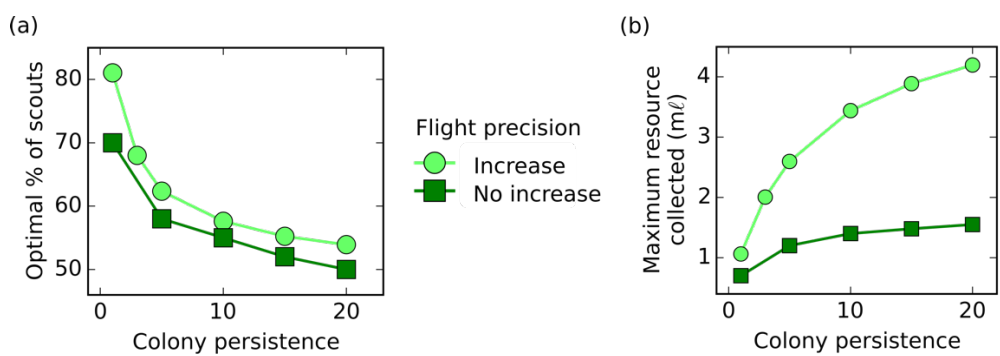

**Figure S6. The effect of increasing flight precision after finding a new resource. (a)** the relationship between optimal colony composition and behavioral persistence was not affected by increasing (light green) flight precision after finding a resource. **(b)** Total amount of resources collected with an increase in flight precision (light green) was on average 2.5 larger than without this increase (dark green). Lines represent the exponential fit to the simulated data.

# Colony size

Regardless of colony size, the proportion of scouts required for collecting the maximum amount of resources (optimal composition) decreased with persistence (Figure S2a). In the main text, we show results from simulations with 300 scouts. The exponential decays for all colony sizes were very close to 40 ± 5 (% scouts/persistence). Although colonies of different sizes showed slightly different optimal compositions, these differences saturated above 900 bees (Figure S2b). Regardless, larger colonies always collected more resources than small ones when given the same amount of time to forage (Figure S2c). For all colony sizes, the maximum amount of resources collected can be approximated by a saturating exponential model, $A_0(1 - e^{\alpha \pi})$, with $\pi$ being persistence (see lines in Figure S2c).

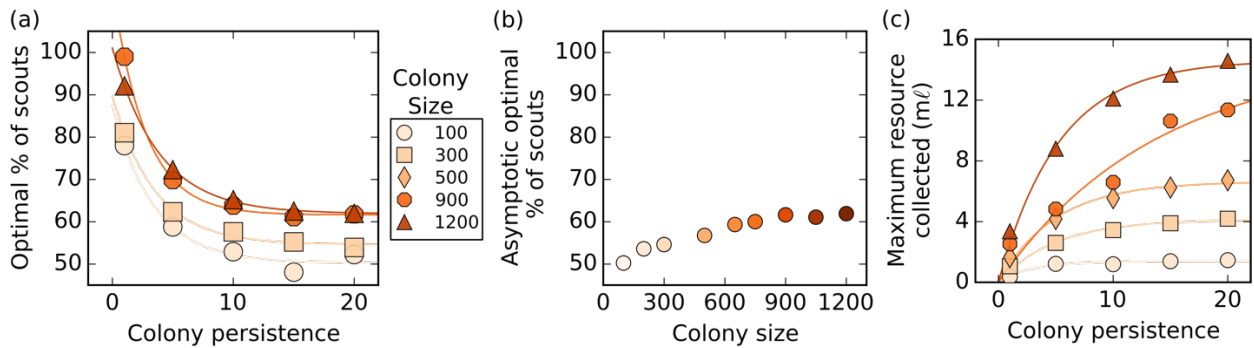

**Figure S2. The size of a colony influences the total amount of resources collected during the simulated foraging, but has little impact on the colony composition that optimizes the collection of resources. (a)** For all colony sizes tested, the optimal percent of scouts decreased with persistence. Lines are the fit of an exponential decay to the simulation results. **(b)** Optimal percent of scouts at the asymptote of the exponential decay changes about 10% with colony size and then stablizes. **(c)** Maximum amount of resources collected increased with colony size. Lines are the fit of an exponential model to the simulaiton results.

# Increasing the rate of recruitment

Increasing the rate of recruitment decreased the proportion of scouts that resulted in the maximum amount of food collected (Figure S3a). Recruitment rate was measured as the average number, K, of bees that were recruited by each waggle dance. When K was greater than 40, the optimal proportion of scouts became very small (below 10% for all persistences). Although the values of the optimal proportion of scouts became smaller, the relationship with colony persistence remained unchanged compared to what we report in the main text. At K=25 the optimal proportion of scouts ranged between 20% and 40% (Figure S3a), in agreement with the estimated percentage of scouts in honeybee colonies [2]. The relationship between the optimal proportion of scouts and the maximum amount of resources collected as a function of the colony persistence remained the same as reported in the main text (Figures S3b,c).

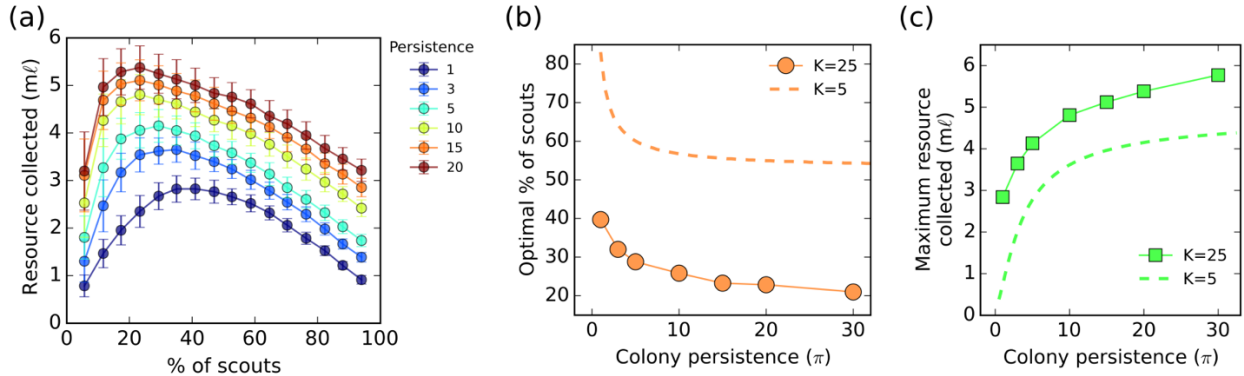

**Figure S3. Effects of increasing the rate of recruitment by scouts.** Recruitment rate is defined as the number, K, of bees recruited by each waggle dance. We compare K = 5 to K=25, which we used in the main text. **(a)** Total amount of resource collected as a function of the proportion of scouts in the colony for different values of persistence. Bars represent the standard deviation across all simulation runs. **(b)** The optimal proportion of scouts ranges from 24% to 40%, considerably smaller than those shown in the main text (dashed line corresponds to Figure 3d). **(c)** Maximum amount of resources collected scales sublinearly with $\pi$. Dashed line represents the results from Figure 3(e).

# Recruitment by recruits

Our simulations include only recruitment by scouts, here we note the effects of adding recruitment by recruits to our model. Recruitment by recruits has the same impact on collective foraging as increasing the rate of recruitment by scouts (see above). To avoid instantaneous depletion of available recruits in the hive, the probability that a recruit will recruit new foragers once it returned to the hive ranged from 0 to 0.5. Adding recruitment by recruits does not change the relationship between the proportion of scouts that results in the maximum amount of food collected and the persistence of foragers in returning to the same food source (Figure S4a). However, the optimal proportions of scouts are smaller than those reported in the main text, when recruits do not recruit (Figure S4b) and the amounts of resources collected are larger (Figure S4c). Furthermore, adding recruitment by recruits does not change the opposite relationship between the optimal proportion of scouts and the persistence of scouts vs. the persistence of recruits that we report in the main text (Figure S4d,e). When the persistence of recruits is fixed, the optimal proportion of scouts grows almost linearly with the persistence of scouts (Figure S4f); when the persistence of scouts is fixed, the optimal proportion of scouts decays with the persistence of recruits and plateaus between 20% and 40% (Figure S4g).

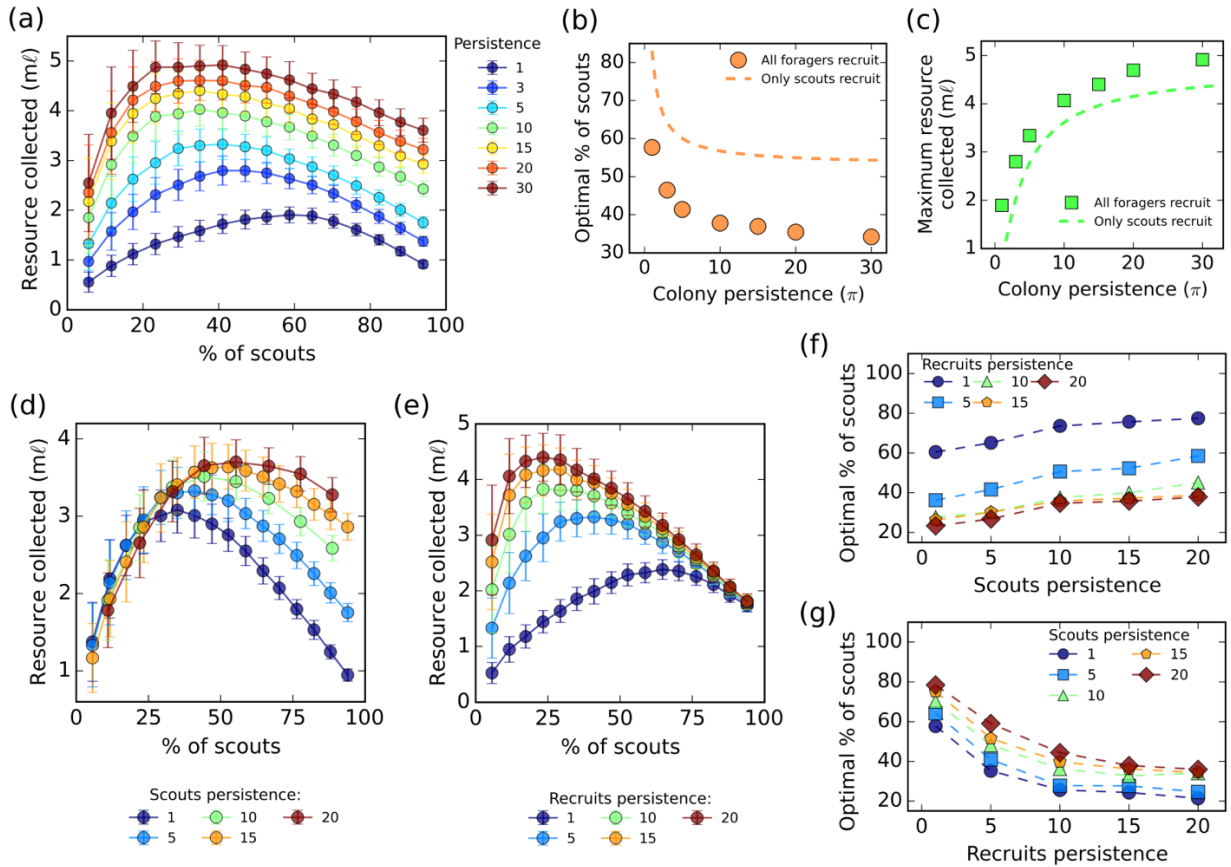

**Figure S4. Effects of including recruitment by recruits in our model.** The relationship between persistence and the optimal proportion of scouts does not change. **(a)** Total amount of resource collected throughout the entire simulation as a function of the proportion of scouts in the colony for different values of persistence of all foragers ($\pi$). Bars represent the standard deviation across all simulation runs. Compare with Figure 3(c). **(b)** The optimal proportion of scouts plateaus near 35% as $\pi$ increases. Dashed line represents the results from Figure 3(d). **(c)** Maximum amount of resources collected scales sublinearly with $\pi$. Dashed line represents the results from Figure 3(e). **(d,e)** Total amount of resources collected by a colony as a function of the proportion of scouts when (d) the persistence of scouts is set to $\pi^s = 5$ for the following values of persistence for recruits: $\pi^r$ =1,5,10,15,20; and (e) the persistence of recruits is set to $\pi^r = 5$ for the following values of persistence of scouts: $\pi^s$ =1,5,10,15,20. Compare with Figures 4(a,b). **(f)** Optimal proportion of scouts as a function of recruit persistence for different values of fixed scout persistence $\pi^s$. Compare with Figure 4(c). **(g)** Optimal proportion of scouts as a function of scout persistence for different values of fixed scout persistence $\pi^r$. Compare with Figure 4(d).

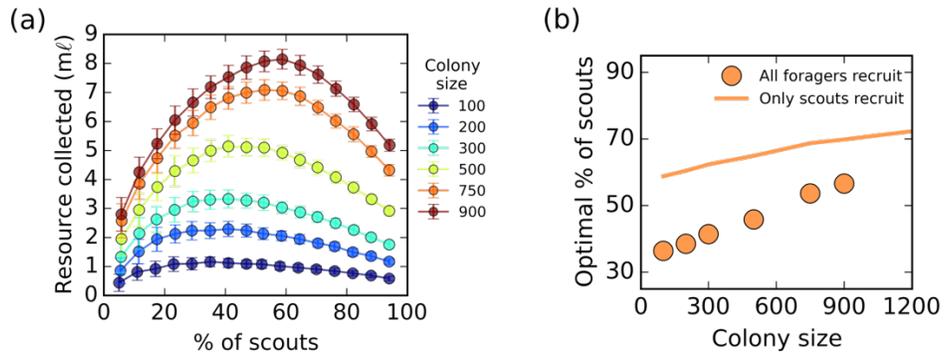

**Figure S5. Relationship between recruitment by recruits and colony size. (a)** Total amount of resource collected as a function of the proportion of scouts in the colony for different colony sizes. Bars represent the standard deviation across all simulation runs. **(b)** The optimal proportion of scouts grows linearly with colony size. The rate at which the proportion of scouts change is faster than that of the results without recruitment by recruits. Compare with Figure S2(a). Persistence for all foragers in both panels was set to $\pi = 5$.

# Number of resource patches

Changing the number of resource patches did not affect our main findings. For each persistence, there was always a group composition that maximized the amount of resources collected (see Figure S3a for one patch). Models with fewer patches had a greater variance in the total amount of resources collected (vertical bars in Figure S3a). Although the optimal composition decreased with persistence, this decrease was sharper for environments with more patches (Figure S3b). Finally, as expected, environments with more patches increased the total amount of resources collected (Figure S3c). The maximum amount of resources collected can be approximated by a saturating exponential model, $A_0 - A_1 e^{\alpha \pi}$, with $\pi$ being persistence (see lines in Figure S3c).

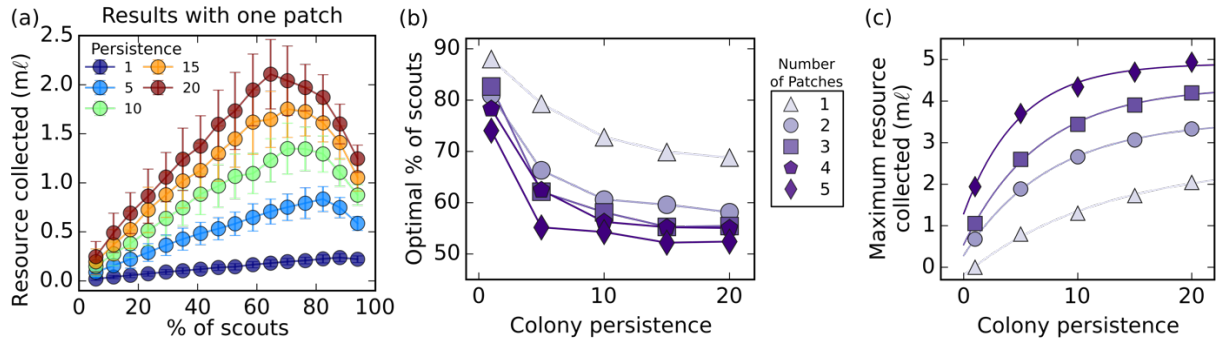

**Figure S6. Changing the number of resource patches in the environment did not affect the relationship between optimal colony composition and persistence. (a)** Amount of resources collected as a function of the percent of scouts when all resources were in a single patch. **(b)** With more patches, the optimal proportion of scouts decreased faster with persistence. **(c)** Maximum amount of resources collected always increased with the number of patches.

# Comparing Predictions between the System Dynamics and Agent-Based Models

The curvature of the dependence between the amount of resources collected and the proportion of scouts slightly differs between the Agent-Based and the Systems Dynamics models (Figure S4). These differences emerge from the fact that the only non-linearity in our Systems Dynamics model is the recruitment term $-S \times R$ (Equation 5b), resulting in curves that are parabolas, as predicted by Equation 10. However, the Agent-Based Model is comprised by many non-linear interactions and processes that are difficult to express analytically. For instance, recruitment was modeled as a stochastic contact process that lasts for a variable period of time; the spatially-explicit distribution of resources slightly deviates from a uniform distribution, which in the systems dynamics model is assumed completely uniform by the fixed the rate $\gamma_d$ at which scouts find new resources; and the flight dynamics in the Agent-Based model is piece-wise continuous, with frequent changes to its functional form (i.e., flight dynamics change from a drifting random walk to movement in a straight line and then to stopping at the hive). Each of these processes change the shape and curvature of the simulated relationship between the amount of resources collected and the proportion of scouts.

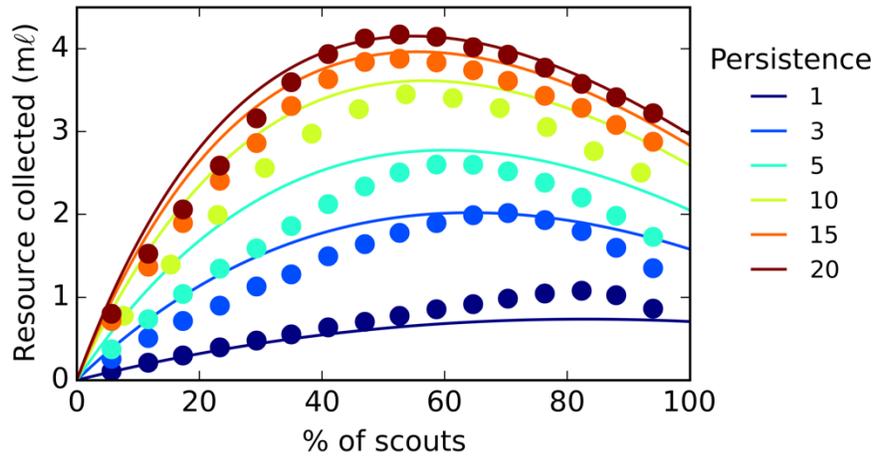

**Figure S7.** Comparing the predicted curves for food collection in relation to the proportion of scouts between the Systems Dynamics model (Equations 5a, 5b and 6, lines) and the Agent-Based model (points).